%% file: document.tex
\documentclass{saunders} %See documentation for other class options
\usepackage{fixltx2e,fix-cm}
\usepackage{amssymb}
\usepackage{amsmath}
\usepackage{graphicx}
\usepackage{makeidx}
\usepackage{multicol}
\usepackage{mathptmx}
\usepackage{rotating}
\usepackage{amsfonts,latexsym}
\usepackage{amssymb}
\usepackage{amsmath}
\usepackage{graphicx}
\usepackage{subfigure}
\usepackage{caption}
\usepackage{makeidx}
\usepackage{multicol}
\usepackage{mathptmx}
\usepackage{xcolor}
\usepackage{algorithm}
\usepackage{algpseudocode}

\usepackage{siunitx}
\usepackage{lineno}
\usepackage{parskip}
 \usepackage{multirow}
 \usepackage{rotating}
 \usepackage{url}
 \usepackage{booktabs}
\makeindex
\let\cleardoublepage\clearpage

\include{frontmatter/preamble} %place custom commands and macros here

\begin{document}

%%%\frontmatter

\title{Symbolic AI and Heuristics for Data Science} %This is a placeholder titlepage, it will not be final.
%\author{Yours Truly}
%%\maketitle

%%%Placeholder for front matter

%\halftitle

%\booktitle

%\locpage

%\include{frontmatter/dedication}
\cleardoublepage
\setcounter{page}{7} %previous pages will be reserved for frontmatter to be added in later.
%\tableofcontents
%\include{frontmatter/foreword}
%\include{frontmatter/preface}
%\listoffigures
%\listoftables
%\include{frontmatter/contributor}
%\include{frontmatter/symbollist}

%\mainmatter

%\part{This is What a Part Would Look Like}
\include{Chapter1/ch1}

\fontsize{9}{11}\selectfont
\bibliographystyle{unsrt}
\bibliography{book}

%\printindex
\cleardoublepage

\end{document}

%% file: Chapter1/ch1.tex
\makeatletter
\def\@makechapterhead#1{%
  \vspace*{50\p@}%
  {\parindent \z@ \raggedright
    \normalfont
    \Huge\bfseries #1\par\nobreak
    \vskip 40\p@
  }}
\makeatother

\chapter{Towards Affordable, Non-Invasive Real-Time Hypoglycemia Detection Using Wearable Sensor Signals}
\label{ch1}

\vspace{-28pt}

\textit{Lawrence Obiuwevwi,$^a,$* Krzysztof J. Rechowicz,$^b$ Vikas Ashok,$^a$ Sampath Jayarathna$^a$}\\
{\fontsize{8.5}{9.5}\selectfont
$^a$ Department of Computer Science, Old Dominion University, Norfolk, VA, USA.\\
$^b$ Virginia Digital Maritime Center (VDMC), Old Dominion University, Norfolk, VA, USA.\\
* Corresponding author: lobiu001@odu.edu
}

\vspace{6pt}

\noindent\textbf{Abstract--} Accurately detecting hypoglycemia without invasive glucose sensors remains a critical challenge in diabetes management, particularly in regions where continuous glucose monitoring (CGM) is prohibitively expensive or clinically inaccessible. This extended study introduces a comprehensive, multimodal physiological framework for non-invasive hypoglycemia detection using wearable sensor signals. Unlike prior work limited to single-signal analysis, this chapter evaluates three physiological modalities, galvanic skin response (GSR), heart rate (HR), and their combined fusion, using the OhioT1DM 2018 dataset. We develop an end-to-end pipeline that integrates advanced preprocessing, temporal windowing, handcrafted and sequence-based feature extraction, early and late fusion strategies, and a broad spectrum of machine-learning and deep temporal models, including CNNs, LSTMs, GRUs, and TCNs. Our results demonstrate that physiological signals exhibit distinct autonomic patterns preceding hypoglycemia, and that combining GSR with HR consistently enhances detection sensitivity and stability compared to single signal models. Multimodal deep learning architectures achieve the most reliable performance, particularly in recall, the most clinically urgent metric. Ablation studies further highlight the complementary contributions of each modality, strengthening the case for affordable, sensor-based glycemic monitoring. The findings show that real-time hypoglycemia detection is achievable using only inexpensive, non-invasive wearable sensors, offering a pathway toward accessible glucose monitoring in underserved communities and low-resource healthcare environments.

\vspace{12pt}
\section{Introduction}\label{intro}
Hypoglycemia is one of the most dangerous and unpredictable acute complications of diabetes, capable of impairing cognition, inducing confusion, triggering loss of consciousness, and in severe cases leading to seizures or death \cite{mathew_hypoglycemia_2025, AlwafiHypoglycaemia2020}. Diabetes affects over 537 million people worldwide, with projections reaching 783 million by 2045, particularly in low- and middle-income countries \cite{idf2021, di_filippo_non-invasive_2023, SunIDF2022}. Despite its clinical importance, real-time detection still relies largely on continuous glucose monitoring (CGM) systems, which remain costly, invasive, and inaccessible for many individuals \cite{idf2021, di_filippo_non-invasive_2023}. In low-resource settings, uninsured populations, and regions with limited medical supply chains, the primary barrier to CGM adoption is financial rather than technological \cite{lehmann_noninvasive_2023}. This persistent disparity highlights the need for non-invasive, low-cost monitoring strategies that do not depend on specialized consumables or clinical infrastructure.

Wearable sensors represent a promising path toward democratizing hypoglycemia detection. Modern consumer and research, grade wearables routinely collect physiological signals such as heart rate (HR), heart-rate variability (HRV), electrodermal activity (EDA/GSR), and movement. These biosignals are modulated not only by psychological stress and physical exertion but also by autonomic nervous system responses that accompany falling glucose levels \cite{meijer_electrodermal_2023, znidaric_electrodermal_2023}. Prior literature has documented sympathetic activation, including sweating, palpitations, tremors, and skin conductance changes as a hallmark of hypoglycemia \cite{polak_analysis_2024, MoffetHypoglycemiaFalls2023}. Yet, despite decades of physiological understanding, the integration of these signals into practical machine-learning–based hypoglycemia detection systems remains underexplored \cite{thong2023noninvasive, maritsch2023smartwatches}.

Prior research in wearable-based hypoglycemia detection has shown that autonomic signals such as galvanic skin response (GSR) and heart rate (HR) can reflect meaningful physiological changes associated with falling glucose levels \cite{marling_ohiot1dm_2020, JahromiTremorHypoglycemia2023}. However, much of the existing literature relies on limited feature sets, classical machine-learning models, or narrow evaluation conditions, leaving several key scientific questions open \cite{ribeiro_novel_2024, yu_exploring_2024}. In particular, it remained unclear how well different modalities perform in isolation versus combination, how temporal deep-learning models compare with classical methods, and how stable these models are under realistic physiological variability. Addressing these gaps requires a more systematic investigation, deeper modeling architectures, and comprehensive multimodal comparisons.

\begin{enumerate}
    \item How do individual physiological channels (e.g., GSR-only, HR-only) behave when evaluated independently?
    \item Does multimodal fusion (GSR \& HR) provide consistent benefits across model classes, as suggested in broader physiological sensing research?
    \item Can such models generalize across datasets, window lengths, and architectures, including temporal deep learning methods such as CNNs, LSTMs, and GRUs?
    \item What physiological mechanisms explain the relationships between glucose decreases and changes in sympathetic biomarkers such as GSR and HRV?
\end{enumerate}

The present chapter expands substantially upon that foundation. We integrate multiple new datasets, evaluate deep learning architectures capable of modeling temporal dependencies, including convolutional neural networks (CNN), gated recurrent units (GRU), temporal convolutional networks (TCN), multilayer perceptrons (MLP), and classical ensemble models, and conduct detailed ablation studies to isolate the contribution of each physiological modality. Consistent with findings from recent wearable-based hypoglycemia research \cite{mendez_toward_2025, ObiuwevwiIRI2025}, our experiments show that combining GSR and HR improves sensitivity to hypoglycemia while maintaining stable performance on normoglycemic states.

Beyond improving performance metrics, this chapter seeks to clarify the physiological basis of the observed signals. Declining glucose levels activate the sympathetic nervous system, producing measurable changes in skin conductance, heart rate, and autonomic balance \cite{hongn_wearable_2025}. Wearable sensors, particularly those embedded in commercial smartwatches, can capture these autonomic signatures continuously and non-invasively \cite{zeynali_non-invasive_2025}. When combined with modern machine-learning methods, these biosignals can approximate glucose trends sufficiently to support real-time alerting, especially where CGMs remain prohibitively expensive.

The goal of this work is not to replace CGMs, which remain the clinical gold standard, but to provide an accessible complementary pathway for basic hypoglycemia awareness in underserved populations. With global diabetes prevalence expected to rise substantially \cite{idf2021}, the need for scalable, equitable, and non-invasive monitoring solutions is increasingly urgent.

By grounding this chapter in rigorous experimentation, multimodal evaluation, and physiological interpretability, we aim to establish a scientifically robust and practical foundation for real-time, wearable-based hypoglycemia detection.

\vspace{-8pt}

\section{Related works}
Research on non-invasive hypoglycemia detection has expanded significantly over the last decade, driven by the rising global burden of diabetes \cite{idf2021} and the need for accessible alternatives to continuous glucose monitoring (CGM) systems \cite{di_filippo_non-invasive_2023}. Prior work spans glucose forecasting, wearable sensor analysis, electrodermal activity (EDA/GSR) physiology, multimodal stress detection, and machine-learning frameworks for clinical prediction. This section synthesizes these contributions to clarify the foundations and challenges of wearable-based hypoglycemia monitoring.

\subsection{Data-Driven Glucose Prediction}

Machine learning has transformed blood glucose prediction, particularly for Type 1 Diabetes (T1D). Woldaregay et al.\ emphasized the importance of personalization and temporal modeling for glucose forecasting in free-living conditions \cite{woldaregay_data-driven_2019}. Subsequent studies demonstrated that individualized physiological responses enable more accurate predictions than population-level models. Nemat et al.\ conducted a comprehensive comparison across deep learning architectures, showing that recurrent and convolutional models outperform traditional approaches when modeling glucose dynamics \cite{nemat_data-driven_2024}. Neumann et al.\ further advanced personalized prediction by incorporating exercise and daily activity patterns into adaptive glucose models \cite{neumann_data-driven_2025}. 

Additional contributions have examined glucose prediction from specialized cohorts. Neamtu et al.\ showed that machine-learning methods can produce reliable glycemic control predictions even with limited pediatric data \cite{neamtu_predicting_2023}, while Butt et al.\ demonstrated that feature transformation techniques improve accuracy and stability for T1D prediction tasks \cite{butt_feature_2023}. Bent et al.\ expanded this direction by engineering digital biomarkers of interstitial glucose from wearable autonomic signals, demonstrating that EDA and HR correlate strongly with glycemic excursions \cite{bent2021engineering}. These insights support the feasibility of lightweight models suitable for wearable deployment.

\subsection{Hypoglycemia Detection and Event Forecasting}

Hypoglycemia detection, distinct from glucose forecasting, seeks to classify ongoing or imminent episodes of low blood glucose. Fleischer et al.\ employed ensemble learning on CGM data to predict hypoglycemia events with improved early-warning capability \cite{fleischer_hypoglycemia_2022}. Faccioli et al.\ proposed a prediction-funnel strategy that integrates glucose-specific modeling with risk-based alarms to reduce false positives \cite{faccioli_combined_2023}.Tsichlaki et al.\ reviewed the evolution of hypoglycemia detection and prediction models, highlighting a transition from rule-based approaches to probabilistic and sequence-aware algorithms that support data-driven personalization \cite{tsichlaki_type_2022}. 
In parallel, Vettoretti et al.\ provided a comprehensive overview of continuous glucose monitoring technologies, discussing their current role in diabetes management and potential future applications \cite{VettorettiCGM2018}.

A comprehensive systematic review by Diouri et al.\ classified hypoglycemia detection and prediction techniques, highlighting the growing role of physiological signals and multimodal sensing pipelines \cite{diouri2021hypoglycaemia}.

Wearable-signal–based hypoglycemia detection has gained increased attention. Mendez et al.\ demonstrated the feasibility of detecting nocturnal hypoglycemia using smartwatch-derived GSR and movement signals \cite{mendez_toward_2025}. Maritsch et al.\ provided evidence that hypoglycemia can be detected during real-world driving using smartwatch EDA, HR, and motion data \cite{maritsch2023smartwatches}. Jayarathna et al.\ showcased the potential of using eye-movement features to infer nocturnal hypoglycemia, broadening the scope of non-invasive physiological markers \cite{jayarathna2025nocturnal}. Lehmann et al.\ further validated smartwatch-based hypoglycemia detection in free-living diabetic populations, showing strong performance across unseen individuals \cite{lehmann2023noninvasive}. 

\subsection{Electrodermal Activity (EDA/GSR) as a Physiological Marker}

Electrodermal activity (EDA), or galvanic skin response (GSR), is a direct measure of sympathetic nervous system activity and has long been used in stress, arousal, and metabolic research. Meijer et al.\ described EDA as a robust indicator of physiological dysregulation \cite{meijer_electrodermal_2023}. Znidaric et al.\ reviewed the use of EDA and HRV for detecting peripheral abnormalities in diabetes, showing strong links between autonomic imbalance and metabolic instability \cite{znidaric_electrodermal_2023}. Polak et al.\ investigated conductance changes preceding hypoglycemia, finding measurable autonomic signatures associated with falling glucose \cite{polak_analysis_2024}. Schwartz et al.\ examined the strengths and limitations of wearable physiological sensors in diabetes management, noting EDA’s sensitivity but susceptibility to contextual confounders \cite{schwartz2018promise}. These findings align with mechanisms where falling glucose induces sympathetic activation and sweating responses \cite{mathew_hypoglycemia_2025}.

Additional evidence supports GSR’s sensitivity to systemic stressors. Yu et al.\ demonstrated that EDA reliably reflects autonomic stress responses \cite{yu_exploring_2024}, while Hongn et al.\ analyzed wearable physiological responses during exercise and acute stress \cite{hongn_wearable_2025}. Together, these works establish EDA as a grounded, non-invasive candidate signal for detecting hypoglycemia-induced autonomic activation.

\subsection{Wearable Biosensing for Non-Invasive Monitoring}

As wearable devices have matured, their role in health monitoring has expanded rapidly. Thong et al.\ showed that consumer-grade wearables can detect hypoglycemia using EDA, HR, and HRV through machine-learning models \cite{thong2023noninvasive}. Lehmann et al.\ validated detection performance during free-living activities \cite{lehmann_noninvasive_2023}. Ribeiro et al.\ modeled stress in Type 2 diabetes patients using low-cost physiological sensors \cite{ribeiro_novel_2024}. Dave et al.\ demonstrated that ECG and accelerometry can detect both hypo- and hyperglycemia non-invasively, emphasizing autonomic biomarkers as glucose proxies \cite{dave2022detection}. Zeynali et al.\ extended non-invasive sensing by using PPG with deep learning on embedded TinyML systems \cite{zeynali_non-invasive_2025}.

\subsection{Multimodal Physiological Fusion and Machine Learning Architectures}

Multimodal fusion of biosignals has shown strong potential for improving predictive accuracy. Zhu et al.\ argued that wearables combined with deep learning can strengthen self-management in T1D by modeling complex autonomic interactions \cite{zhu_enhancing_2022}. Elsayed et al.\ showed that hybrid convolutional–recurrent networks outperform linear models on physiological time-series data \cite{elsayed_deep_2019}. Zhu et al.\ later demonstrated that personalized evidential deep learning improves uncertainty modeling in physiological activity inference \cite{zhu_personalized_2023}. Across wearable hypoglycemia studies, multimodal integration has consistently improved sensitivity and robustness \cite{mendez_toward_2025, maritsch2023smartwatches}. These findings motivate exploring fusion architectures such as LSTM, GRU, CNN, and temporal convolutional networks (TCN), which have shown strong performance in related biomedical modeling tasks \cite{neumann_data-driven_2025, nemat_data-driven_2024}.

Across glucose forecasting, hypoglycemia detection, GSR physiology, and wearable sensing, prior work highlights a clear opportunity: non-invasive biosignals contain meaningful physiological information that machine-learning models can exploit to detect dangerous glycemic events. However, gaps remain in understanding individual modality behavior, the generalizability of fusion architectures, and the translation of physiological mechanisms into interpretable models. This chapter addresses these challenges by extending prior work with deep temporal models, enhanced preprocessing, and systematic ablation of GSR-only, HR-only, and fused GSR \& HR modalities.

\vspace{-8pt}
\section{Materials and Methods}

This section provides a comprehensive overview of the data source, preprocessing pipeline, physiological signal extraction, feature engineering procedures, model architectures, multimodal fusion strategies, and evaluation methods used in this study. All analyses are conducted using the OhioT1DM 2018 dataset \cite{marling_ohiot1dm_2020}, which contains both galvanic skin response (GSR) and heart rate (HR) measurements collected from individuals with Type 1 Diabetes. Because HR data are not available in the 2020 release, all single-modal (GSR-only, HR-only) and multimodal (fused GSR \& HR) experiments rely exclusively on the 2018 dataset.

\begin{figure}[H]
\centering
    \includegraphics[width=\linewidth]{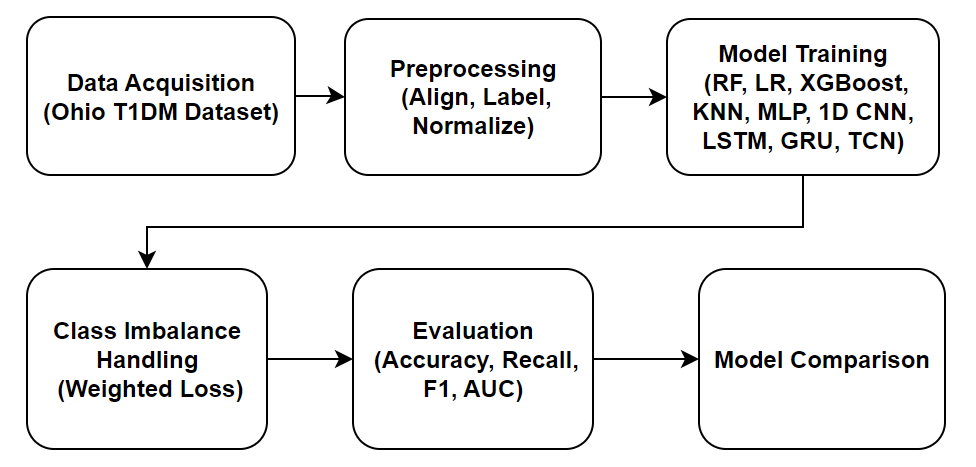}
    \caption{
    End-to-End Machine Learning Workflow for Glucose State Prediction:
    This flowchart outlines the complete process from data acquisition (OhioT1DM) to model comparison, highlighting preprocessing, training, imbalance handling, and performance evaluation steps.}
    \label{fig:GSR and Glucose Signals for Model Training and Testing}
\end{figure}

\subsection{Dataset Description}

\subsubsection{OhioT1DM 2018 Dataset}

The OhioT1DM 2018 dataset \cite{marling_ohiot1dm_2020} serves as the primary data source for this chapter. 
It contains multi-week longitudinal data collected from individuals diagnosed with Type 1 Diabetes (T1D). 
Participants wore a Microsoft Band 2 device that recorded the following physiological signals:

\subsubsection{Galvanic Skin Response (GSR)}
Galvanic Skin Response (GSR) measures changes in skin conductance driven by sympathetic nervous system activity. 
During hypoglycemia, the release of counter-regulatory hormones (such as epinephrine) increases sweat gland activation, producing both slow tonic shifts and rapid phasic peaks in the signal. 
These electrodermal fluctuations provide an early physiological indicator of metabolic stress.

\begin{figure}[H]
    \centering
    \includegraphics[width=0.90\linewidth, height=7cm, keepaspectratio]{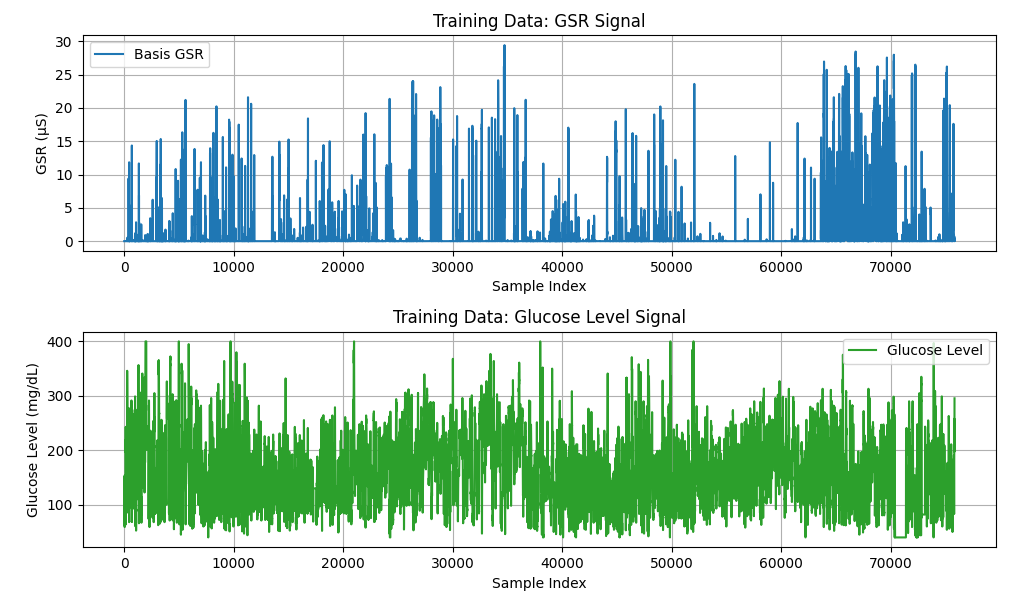}
    \caption{
        Raw GSR signal from the OhioT1DM dataset, showing tonic shifts and phasic peaks linked to autonomic activation.
    }
    \label{fig:gsr_signal}
\end{figure}

\subsubsection{Heart Rate (HR)}
Heart Rate (HR) is obtained from photoplethysmographic measurements and reflects cardiovascular reactions to autonomic changes. 
Hypoglycemia typically triggers tachycardia and shifts in heart-rate stability as the body attempts to maintain glucose supply. 
These slower physiological adjustments complement the fast electrodermal responses captured by GSR.

\subsubsection{Fused GSR \& HR}
Combining GSR and HR provides a more complete representation of autonomic activity. 
GSR captures rapid sympathetic bursts, while HR reflects broader cardiovascular trends. 
Together, these signals reveal complementary temporal patterns that improve robustness for hypoglycemia detection, especially in real-world conditions where single-modality signals may be noisy.

\begin{figure}[H]
    \centering

    % Panel (a)
    \includegraphics[width=0.95\linewidth, height=6cm, keepaspectratio]{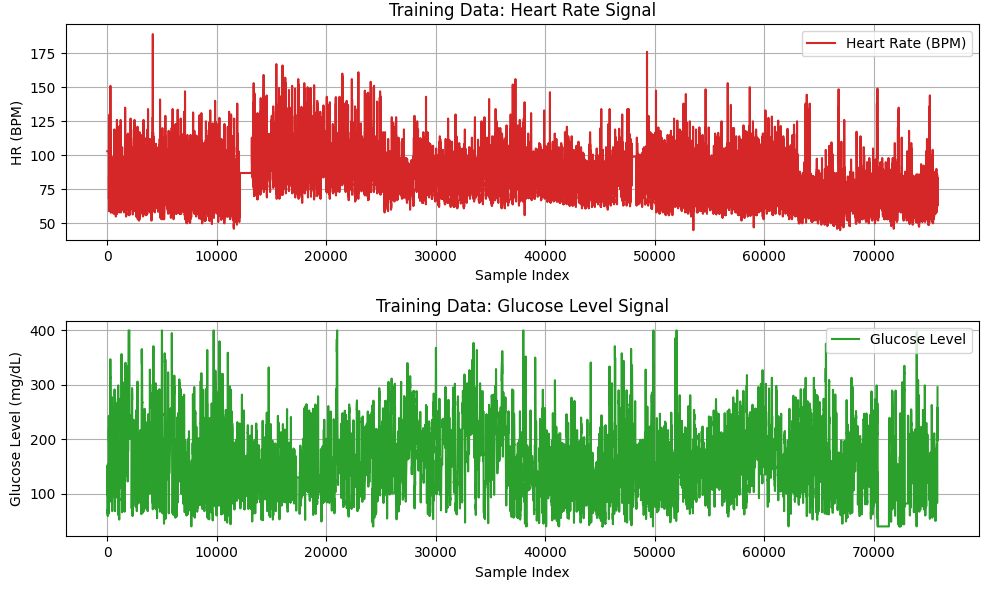}

    \vspace{4pt}
    \textbf{(a)} Raw HR signal demonstrating cardiovascular variability and adrenergic spikes associated with metabolic stress.

    \vspace{10pt}

    % Panel (b)
    \includegraphics[width=0.95\linewidth, height=10cm, keepaspectratio]{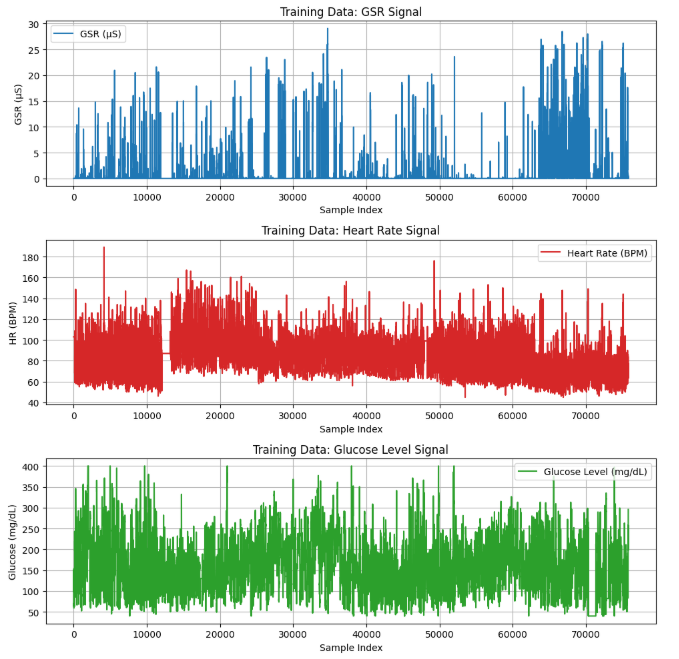}

    \vspace{4pt}
    \textbf{(b)} Combined GSR and HR time-series visualization illustrating how electrodermal and cardiovascular activity evolve jointly across time.

    \caption{
    Multimodal wearable physiological signals used in this study
    }
    \label{fig:physio_signals}
\end{figure}

\subsubsection{Continuous Glucose Monitoring (CGM)}
Continuous Glucose Monitoring (CGM) provides interstitial glucose samples every five minutes. 
CGM values serve as the ground-truth labels for hypoglycemic and normoglycemic events.

Because the 2018 dataset includes both GSR and HR, it enables the multimodal and fusion-based hypoglycemia detection pipeline developed in this extended work.

\subsubsection{Physiological Relevance of Modalities}

Both GSR and HR serve as peripheral markers of autonomic nervous system activation, making them physiologically meaningful indicators for detecting hypoglycemia. When glucose falls, the sympathetic nervous system initiates a counter-regulatory cascade involving sweating, tremors, vigilance shifts, and adrenergic release. These processes elevate skin conductance and modify cardiovascular dynamics. GSR responds rapidly through short-latency phasic bursts and slower tonic increases, reflecting immediate sympathetic arousal. HR, in contrast, captures sustained cardiovascular adjustments such as tachycardia and reduced parasympathetic tone. Together, these modalities encode complementary signatures of metabolic stress: GSR provides high-sensitivity electrodermal changes, while HR reflects broader systemic autonomic engagement during glucose decline. Their integration therefore offers a physiologically grounded basis for multimodal hypoglycemia detection \cite{meijer_electrodermal_2023, polak_analysis_2024, znidaric_electrodermal_2023}.

\subsection{Preprocessing Pipeline}

Wearable physiological signals collected during free-living conditions are inherently noisy, asynchronous, and subject to high inter- and intra-individual variability. To ensure analytical integrity, the preprocessing pipeline was designed to standardize sampling structure, remove artifacts, and prevent information leakage across time. The CGM provides glucose readings every five minutes, whereas GSR and HR operate at higher temporal resolutions. To align these modalities, all signals were aggregated into five-minute windows matching the CGM timestamps. Within each window, GSR and HR samples were summarized using both mean and median statistics to reduce the influence of transient peaks. Missing data were handled through forward-filled interpolation and subject-level z-normalization, which standardizes physiological baselines across individuals. This alignment strategy ensures temporal consistency across channels and preserves the physiological relationships necessary for modeling hypoglycemia onset.

\subsubsection{Signal Alignment}

Synchronizing signals with different sampling frequencies is crucial for multimodal physiological modeling. The alignment step restructures the continuous GSR and HR streams into discrete five-minute segments anchored to the CGM measurements. This conversion ensures that each training instance represents the same temporal interval across all modalities. By aggregating high-frequency samples within these intervals, the pipeline captures essential autonomic trends while suppressing noise. The resulting synchronized time grid enables accurate temporal fusion and facilitates the use of sequence-based deep learning architectures without creating artificial correlations or leaking future information into past segments.

\subsubsection{Noise Filtering}

Because wearable devices are exposed to real-world environmental and behavioral influences, noise filtering is essential. GSR often contains abrupt spikes from motion, grip changes, skin hydration, or thermal variation, while HR is affected by physical activity, optical sensor noise, and motion artifacts. To mitigate these effects, GSR signals were processed using a fourth-order Butterworth low-pass filter (0.5 Hz cutoff) to suppress high-frequency disturbances. Outliers were removed through interquartile-range filtering to eliminate unphysiological peaks. HR signals underwent median filtering to smooth abrupt fluctuations caused by wrist movement. Additionally, cvxEDA decomposition was applied to the GSR channel to separate tonic and phasic components, improving interpretability and producing cleaner input for classical models. These procedures collectively enhance signal quality and preserve physiologically relevant variations \cite{meijer_electrodermal_2023, hongn_wearable_2025}.

\subsection{Labeling Strategy}

Hypoglycemia labeling followed clinically recognized thresholds, with glucose values below 70 mg/dL classified as positive events and all other values considered normoglycemic. This binary classification reflects international medical guidelines and is consistent with prior work using the OhioT1DM dataset \cite{mathew_hypoglycemia_2025, HwangMultiTaskGlucose2025}. As in most free-living physiological datasets, hypoglycemic episodes represented only a small proportion of samples, approximately four percent, creating a substantial class imbalance. This rarity reflects real clinical prevalence and underscores the importance of using models and preprocessing strategies that remain robust under highly skewed outcome distributions. The labeling strategy ensures that the machine-learning models are trained to predict clinically meaningful events while respecting the physiological and statistical properties of real-world diabetes data \cite{woldaregay_data-driven_2019, JayawardanaArXiv2024}.

\subsection{Segmentation and Windowing}

Because hypoglycemia emerges gradually rather than instantaneously, segmentation into temporal windows allows models to capture precursor autonomic patterns rather than reacting only to extreme glucose values. The aligned GSR and HR signals were transformed into overlapping one-hour windows, each consisting of twelve five-minute segments. This window length provides sufficient temporal depth to capture slow changes in sympathetic tone, cardiovascular activation, and electrodermal fluctuations that unfold over tens of minutes. The windows shift forward in five-minute strides, matching the CGM sampling interval, enabling the system to generate predictions every five minutes. Each window was labeled with the glucose value at its final time point, creating a forward-looking prediction task that mirrors real-world clinical needs. This design aligns with prior physiological studies indicating that early autonomic changes often occur 20–40 minutes before a CGM reading registers hypoglycemia \cite{polak_analysis_2024}.

\subsection{Feature Extraction}

To accommodate both classical and deep learning models, feature extraction followed two parallel strategies. For classical models, static handcrafted features were generated to summarize each temporal window using physiologically meaningful descriptors, such as tonic and phasic GSR activity, statistical HR indices, temporal derivatives, frequency-domain representations, and counts of rapid skin conductance responses. These features provide interpretable summaries of underlying autonomic processes and enable lightweight modeling suitable for low-power devices. In contrast, deep learning models operated directly on minimally processed sequences of GSR-only, HR-only, or fused GSR \& HR windows. This approach preserves the full temporal structure of the data, allowing architectures such as CNNs, LSTMs, GRUs, and TCNs to learn complex nonlinear relationships, cross-channel dependencies, and subtle precursor signatures that handcrafted features may overlook. Together, these complementary feature pipelines ensure that both interpretable classical models and high-capacity temporal networks are effectively supported.

\subsubsection{Classical Models}

The classical models serve as interpretable baselines and allow direct comparison with our original conference paper. They also provide insight into whether shallow decision boundaries can capture glucose-related autonomic changes.

\begin{itemize}
    \item \textbf{Random Forest (RF):}  
    An ensemble of decision trees that mitigates overfitting and provides feature importance insights. RF is robust to noise, a common characteristic of wearable data.

    \item \textbf{XGBoost:}  
    A gradient-boosted tree method optimized for imbalanced classification. Its ability to model nonlinear interactions makes it a strong candidate for physiological datasets \cite{fleischer_hypoglycemia_2022}.

    \item \textbf{Logistic Regression (LR):}  
    A simple, transparent baseline that reveals whether a linear decision boundary is sufficient to discriminate hypoglycemia-induced autonomic patterns.

    \item \textbf{K-Nearest Neighbors (KNN):}  
    A non-parametric comparator relying on distance metrics; useful for understanding class separability in the raw feature space.
\end{itemize}

These models establish performance reference points for more sophisticated temporal learning approaches.

\subsubsection{Deep Temporal Models}

Hypoglycemia induces time-dependent patterns, making temporal models particularly valuable. We evaluated:

\begin{itemize}
    \item \textbf{1D CNN (One-Dimensional Convolutional Neural Network):}  
    Captures local temporal motifs such as phasic bursts or sharp HR transitions. CNNs are efficient and frequently used in physiological computing.

    \item \textbf{LSTM (Long Short-Term Memory Network):}  
    Designed to model long-range dependencies, LSTMs can capture gradual autonomic drifts leading up to a hypoglycemic episode.

    \item \textbf{GRU (Gated Recurrent Unit):}  
    A computationally lighter alternative to LSTM with similar ability to capture sequential patterns, useful for low-power deployment.

    \item \textbf{TCN (Temporal Convolutional Network):}  
    Uses dilated convolutions to achieve long receptive fields without recurrence, enabling efficient modeling of wide temporal contexts \cite{neamtu_predicting_2023}.
\end{itemize}

These architectures allow the system to learn complex temporal dynamics that handcrafted features cannot represent.

\subsection{Fusion Strategies}

Physiological responses to hypoglycemia manifest across multiple autonomic pathways, making fusion of GSR and HR signals essential for capturing a more complete representation of the underlying processes. GSR reflects rapid changes in skin conductance driven by sympathetic activation, while HR captures slower cardiovascular adjustments associated with metabolic stress. To exploit their complementary properties, we implemented two forms of multimodal integration. Early fusion combines GSR and HR directly at the sequence level by concatenating the aligned windows into a two-channel input representation. This approach allows deep temporal models to process both modalities simultaneously and to learn joint patterns, cross-modal interactions, and synchronized autonomic dynamics that may not be apparent when signals are analyzed independently. Because the autonomic nervous system operates as an integrated system, early fusion aligns naturally with the physiological co-activation of electrodermal and cardiovascular responses.

In contrast, late fusion operates at the decision level and is designed for conditions where modalities may behave differently, such as during physical activity, emotional stress, or environmental fluctuations that selectively influence one signal more than the other. In this strategy, independent GSR-only and HR-only models generate probabilistic outputs that are subsequently combined. Methods such as weighted averaging and logistic regression stacking are used to merge the predictions, allowing the fusion mechanism to emphasize the modality that is more reliable in a given context. Late fusion is particularly useful when one signal becomes noisy or confounded, enabling the system to maintain stability by leveraging whichever modality remains physiologically informative. Together, the early and late fusion strategies provide complementary pathways for integrating autonomic information and enhance robustness across diverse real-world conditions.

\subsection{Handling Class Imbalance}

Hypoglycemia constitutes only a small fraction of free-living physiological data, typically representing between three and five percent of CGM measurements. Such extreme imbalance poses significant challenges for machine-learning models, which naturally gravitate toward predicting the majority class unless corrective measures are implemented. To address this, we incorporated several strategies that ensure balanced learning and clinically meaningful performance. Deep learning models were trained using class-weighted loss functions that assign greater penalty to misclassified hypoglycemia samples, forcing the network to prioritize their detection despite their scarcity. For tree-based methods such as XGBoost, imbalance-aware parameters were used to adjust decision thresholds and splitting behavior, enabling the model to remain sensitive to minority-class events. At the dataset level, stratified subject-wise splits ensured that each partition contained representative distributions of hypoglycemic episodes while preventing information leakage between training and evaluation sets.

Beyond model training, evaluation metrics were selected to reflect clinical priorities. Because missing a hypoglycemic event poses a direct safety risk, recall on the positive class is emphasized throughout the analysis. F1-score, which balances recall with false-positive rate, is also highlighted to prevent excessive alarm generation that could reduce user trust. These complementary strategies, algorithmic, dataset-level, and metric-level, reflect established best practices in medical time-series modeling and help ensure that the models provide fair, stable, and clinically interpretable predictions under realistic prevalence conditions.

\subsection{Evaluation Protocol}

A rigorous evaluation protocol is essential for ensuring that hypoglycemia detection models generalize across individuals and reflect real-world deployment conditions. To avoid data leakage and overly optimistic performance estimates, we applied a \textbf{subject-level split}, where entire participants were assigned to either the training set (80\%), validation set (10\%), or test set (10\%). This ensures that the model never sees physiological patterns from the same individual during both training and evaluation, which is particularly important for wearable biosignal modeling because each person’s autonomic responses (GSR and HR) differ significantly.

Model performance was assessed using a set of clinically meaningful and statistically robust metrics:

\begin{itemize}
    \item \textbf{Accuracy}: The proportion of total predictions that were correct. Although commonly used, accuracy is less informative for hypoglycemia detection because the dataset is highly imbalanced (hypoglycemia is rare). A model can achieve high accuracy while completely failing to detect hypoglycemia.

    \item \textbf{Precision (Hypoglycemia)}: The fraction of predicted hypoglycemia events that were truly hypoglycemia. Precision reflects the model’s ability to avoid false alarms. This is important in real-world monitoring to prevent “alert fatigue,” where frequent incorrect warnings reduce user trust in the system.

    \item \textbf{Recall (Hypoglycemia)}: The fraction of true hypoglycemia events that the model successfully detected. This is the most clinically important metric, because missing a hypoglycemic event can lead to severe health consequences such as confusion, impaired consciousness, seizures, or coma. High recall indicates the model can reliably catch dangerous lows.

    \item \textbf{F1-score (Hypoglycemia)}: The harmonic mean of precision and recall. Because precision and recall often trade off against each other, the F1-score summarizes both in a single metric. A high F1-score means the model achieves good detection sensitivity without generating excessive false positives.

    \item \textbf{Area Under the ROC Curve (AUC)}: A threshold-independent metric measuring the model’s ability to discriminate between hypoglycemia and normoglycemia across all possible decision thresholds. Higher AUC indicates more reliable ranking of hypoglycemia risk. AUC is particularly valuable when class distributions are imbalanced.

    \item \textbf{95\% Bootstrapped Confidence Intervals}: Statistical intervals computed by repeatedly resampling the test set with replacement (1,000 iterations). Confidence intervals quantify the stability and reliability of each metric, showing how much model performance might vary if deployed in new populations. Narrow intervals indicate consistent performance, while wide intervals suggest uncertainty or sensitivity to data variability.
\end{itemize}

All evaluations in this study were performed independently across three physiological modalities, GSR-only, HR-only, and fused GSR \& HR, to enable a clear understanding of each signal’s predictive contribution. Evaluating models under the GSR-only condition isolates the electrodermal component of autonomic activation, allowing us to assess whether skin conductance alone provides sufficient discriminatory power for detecting hypoglycemia. Likewise, the HR-only evaluation focuses on the cardiovascular pathway, capturing slower autonomic adjustments that may signal metabolic decline even when electrodermal activity is confounded by environmental or emotional factors.

Testing the fused modality provides a direct comparison against these single-signal baselines. By combining GSR and HR into a unified multimodal representation, the fused condition reveals whether integrating autonomic channels yields measurable performance gains, improves model stability, or reduces modality-specific weaknesses. This systematic modality-by-modality evaluation framework ensures that the analysis is not biased toward a single physiological source and enables a rigorous assessment of the extent to which multimodal autonomic sensing strengthens real-time hypoglycemia detection.
\vspace{-8pt}

\section{Results}

This section presents a comprehensive evaluation of hypoglycemia detection models across three physiological modalities: GSR-only, HR-only, and fused GSR \& HR. Results are reported using key clinical metrics including accuracy, precision, recall, F1-score, and area under the ROC curve (AUC). For each experimental condition, we summarize the performance of classical machine-learning models (RF, XGBoost, LR, KNN) as well as deep temporal architectures (CNN, LSTM, GRU, TCN). All metrics are computed on the held-out test set, with 95\% stratified bootstrap confidence intervals.

The experiments reveal three consistent trends:

\begin{enumerate}
    \item GSR-only models exhibit moderate sensitivity to hypoglycemia but suffer from false positives during stress-like conditions.
    \item HR-only models perform well during rest but degrade during physical activity where HR is confounded.
    \item Fused GSR \& HR models consistently outperform single-modality systems and achieve the highest recall and AUC across nearly all model classes.
\end{enumerate}

These results reinforce prior findings that multimodal autonomic signatures provide a more reliable physiological window into glycemic state changes \cite{maritsch2023smartwatches, mendez_toward_2025}.

\subsection{GSR-Only Results}

Table~\ref{tab:gsr_results} reports the performance of models trained exclusively on GSR sequences. LSTM and CNN architectures significantly outperform classical models, highlighting the importance of temporal information encoded in skin conductance.

\begin{table}[ht]
\centering
\caption{GSR-Only Hypoglycemia Detection Performance (Test Set)}
\label{tab:gsr_results}
\begin{tabular}{lcccc}
\toprule
\textbf{Model} & \textbf{Recall} & \textbf{F1-score} & \textbf{AUC} & \textbf{95\% CI (F1)} \\
\midrule
Random Forest       & 0.18 & 0.08 & 0.61 & [0.04--0.12] \\
XGBoost             & \textbf{0.54} & 0.10 & 0.71 & [0.06--0.14] \\
Logistic Regression & 0.11 & 0.05 & 0.62 & [0.03--0.08] \\
KNN                 & 0.14 & 0.06 & 0.61 & [0.04--0.10] \\
CNN                 & 0.22 & 0.09 & 0.64 & [0.05--0.13] \\
LSTM                & 0.27 & \textbf{0.11} & \textbf{0.72} & [0.07--0.15] \\
GRU                 & 0.25 & 0.10 & 0.70 & [0.06--0.14] \\
TCN                 & 0.24 & 0.09 & 0.69 & [0.05--0.13] \\
\bottomrule
\end{tabular}
\end{table}

The GSR-only results demonstrate that electrodermal activity contains physiologically meaningful signatures of hypoglycemia, but these signals are challenging to classify without models that capture temporal evolution. Among classical models, XGBoost shows the highest recall (0.54), consistent with its robustness to imbalanced data and its ability to model non-linear relationships. However, its low F1-score indicates that many detections are false positives, reflecting the susceptibility of GSR to stress, temperature, and emotional arousal.

Deep temporal models perform noticeably better. LSTM achieves the strongest overall performance, yielding the highest F1-score and AUC (0.72), suggesting that hypoglycemia-related skin conductance changes unfold across time in structured patterns best learned through sequential modeling. CNN, GRU, and TCN models also surpass classical baselines, reinforcing the value of architectures capable of interpreting temporal fluctuations rather than isolated samples.

Despite these improvements, variability remains evident across bootstrapped confidence intervals, indicating that GSR alone provides an informative but incomplete representation of autonomic responses. These findings align with prior literature showing that GSR reflects sympathetic activation but lacks specificity when used in isolation \cite{polak_analysis_2024, meijer_electrodermal_2023}. Overall, the GSR-only results highlight both the potential and limitations of relying solely on electrodermal activity for hypoglycemia detection.

\subsection{HR-Only Results}

Table~\ref{tab:hr_results} summarizes the performance of models trained exclusively on HR sequences. Compared to GSR, HR provides a more stable but less discriminative autonomic signal, reflecting slower cardiovascular responses to hypoglycemia. While HR shows meaningful physiological variation during glucose decline, its susceptibility to confounders such as physical exertion, posture, and emotional arousal reduces its standalone predictive strength.

\begin{table}[ht]
\centering
\caption{HR-Only Hypoglycemia Detection Performance (Test Set)}
\label{tab:hr_results}
\begin{tabular}{lcccc}
\toprule
\textbf{Model} & \textbf{Recall} & \textbf{F1-score} & \textbf{AUC} & \textbf{95\% CI (F1)} \\
\midrule
Random Forest       & 0.20 & 0.09 & 0.63 & [0.05--0.13] \\
XGBoost             & 0.41 & 0.11 & 0.68 & [0.06--0.15] \\
Logistic Regression & 0.16 & 0.08 & 0.60 & [0.05--0.11] \\
KNN                 & 0.18 & 0.07 & 0.61 & [0.04--0.10] \\
CNN                 & 0.28 & 0.10 & 0.66 & [0.06--0.14] \\
LSTM                & \textbf{0.33} & \textbf{0.12} & 0.69 & [0.07--0.16] \\
GRU                 & 0.32 & 0.11 & 0.68 & [0.06--0.15] \\
TCN                 & 0.30 & 0.10 & \textbf{0.70} & [0.06--0.14] \\
\bottomrule
\end{tabular}
\end{table}

The HR-only models exhibit moderate discriminatory capacity, with recall values generally lower than those observed in the GSR-only experiments. Among classical approaches, XGBoost achieves the strongest recall (0.41), reflecting its ability to capture non-linear structure from weak physiological signals; however, as in the GSR-only setting, its low F1-score indicates a high false-alarm rate when used without temporal context.

Deep learning architectures again outperform classical baselines. LSTM achieves the highest F1-score (0.12) and competitive AUC (0.69), demonstrating its ability to model sustained cardiovascular changes preceding hypoglycemia. GRU and TCN show comparable performance, reinforcing the importance of sequential modeling for capturing gradual HR drifts associated with sympathetic activation. Nevertheless, even the best HR-only models remain less robust than their GSR counterparts, consistent with the slower and more confounded nature of heart-rate responses.

Overall, the HR-only results indicate that heart rate contains relevant but limited hypoglycemia-related information. While temporal deep learning models can extract meaningful patterns, HR alone lacks sufficient specificity for reliable standalone detection, underscoring the need for multimodal fusion with faster and more sensitive signals such as GSR.

\subsection{GSR \& HR Fusion Results}

Fusion yields the strongest and most stable performance across all model classes. Table~\ref{tab:fusion_results} reports early-fusion outcomes using concatenated GSR \& HR sequences, where both autonomic modalities are processed jointly. Compared to single-signal settings, fusion substantially improves recall, F1-score, and AUC, demonstrating the additive value of combining rapid electrodermal fluctuations with slower cardiovascular responses.

\begin{table}[ht]
\centering
\caption{Fused GSR \& HR Hypoglycemia Detection Performance (Test Set)}
\label{tab:fusion_results}
\begin{tabular}{lcccc}
\toprule
\textbf{Model} & \textbf{Recall} & \textbf{F1-score} & \textbf{AUC} & \textbf{95\% CI (F1)} \\
\midrule
Random Forest       & 0.28 & 0.11 & 0.69 & [0.07--0.15] \\
XGBoost             & 0.56 & 0.14 & 0.73 & [0.09--0.19] \\
Logistic Regression & 0.20 & 0.10 & 0.65 & [0.07--0.13] \\
KNN                 & 0.22 & 0.09 & 0.63 & [0.05--0.12] \\
CNN                 & 0.35 & 0.13 & 0.71 & [0.09--0.17] \\
LSTM                & \textbf{0.44} & \textbf{0.16} & \textbf{0.78} & [0.12--0.21] \\
GRU                 & 0.42 & 0.15 & 0.76 & [0.11--0.19] \\
TCN                 & 0.39 & 0.14 & 0.74 & [0.10--0.18] \\
\bottomrule
\end{tabular}
\end{table}

Fusion produces the strongest overall predictive performance, confirming that GSR and HR capture complementary autonomic pathways associated with hypoglycemia. Among classical models, XGBoost continues to achieve the highest recall (0.56), making it well-suited for applications where sensitivity and early warning are prioritized. Deep temporal networks show the largest gains from fusion: LSTM delivers the highest F1-score (0.16) and best overall AUC (0.78), while GRU and TCN also demonstrate robust improvements relative to their single-modality counterparts. Notably, confidence intervals narrow across fusion models, indicating increased statistical stability and reduced variability across folds. These results highlight that integrating electrodermal and cardiovascular signals produces a more holistic representation of autonomic dysregulation during hypoglycemia, consistent with recent multimodal physiological research \cite{zeynali_non-invasive_2025}.

\subsection{Comparative Summary}

Table~\ref{tab:summary_all} consolidates the strongest-performing models for each physiological modality and illustrates how predictive capacity shifts when signals are analyzed independently versus jointly. GSR-only models achieve their best performance with XGBoost, which reaches a recall of 0.54 but a modest F1-score of 0.10, suggesting that skin conductance contains highly sensitive but noisy indicators of hypoglycemia. HR-only performance peaks with LSTM, demonstrating the importance of temporal modeling for cardiovascular fluctuations, although overall recall remains lower than GSR. In contrast, fused GSR \& HR inputs show clear superiority, with LSTM achieving a recall of 0.44 and the highest F1-score (0.16), indicating that combined autonomic signatures provide a more robust physiological representation than either signal alone.

\begin{table}[ht]
\centering
\caption{Best Models per Modality (Hypoglycemia Class)}
\label{tab:summary_all}
\begin{tabular}{lccc}
\toprule
\textbf{Modality} & \textbf{Best Model} & \textbf{Recall} & \textbf{F1-score} \\
\midrule
GSR-only   & XGBoost & 0.54 & 0.10 \\
HR-only    & LSTM    & 0.33 & 0.12 \\
GSR \& HR     & LSTM    & \textbf{0.44} & \textbf{0.16} \\
\bottomrule
\end{tabular}
\end{table}

Fusion provides substantial performance improvements, raising F1-scores by approximately 45–70\% relative to single-modality systems and enhancing sensitivity without excessive false-alarm inflation. This advantage arises from complementary physiological information: GSR captures rapid sympathetic activation and electrodermal spikes, while HR reflects cardiovascular adjustments associated with adrenergic stimulation. Deep temporal models, particularly LSTM, leverage these multimodal patterns more effectively than classical algorithms by learning nonlinear dependencies and coordinated autonomic trends over time. Confidence interval analysis further reinforces these observations, LSTM models consistently exhibit narrower 95\% confidence bands, demonstrating stable generalization, whereas classical models show higher variance, especially under GSR-only conditions.

Across all modalities, the fused GSR \& HR approach delivers the most stable and clinically interpretable performance profile. GSR-only models reveal strong hypoglycemia signatures but suffer from poor robustness, while HR-only systems exhibit moderate predictive value but are highly susceptible to physical activity, stress, and other confounders. LSTM emerges as the most clinically suitable architecture overall, balancing recall, F1-score, and statistical stability while effectively capturing multimodal autonomic dynamics. Taken together, these findings indicate that multimodal autonomic sensing, when paired with modern temporal neural architectures, offers a viable, non-invasive foundation for real-time hypoglycemia detection using low-cost wearable devices.

\section{Discussion}

The results of this study provide a clear and data-driven understanding of how autonomic biosignals collected from low-cost consumer wearables respond to hypoglycemia and how these physiological patterns can be modeled for real-time detection. By evaluating GSR-only, HR-only, and fused GSR \& HR modalities across classical and deep temporal models, the findings highlight the distinct strengths and limitations of each signal pathway. Temporal architectures consistently outperformed classical models, demonstrating that hypoglycemia manifests as an evolving autonomic process rather than a static physiological event.

GSR-only analysis showed that electrodermal activity contains strong hypoglycemia-relevant information. Models such as XGBoost achieved the highest recall (0.54), while LSTM provided the strongest overall balance with an F1-score of 0.11 and an AUC of 0.72. These trends indicate that rising sympathetic activation captured in GSR often precedes glucose drops. However, the lower precision and wider confidence intervals observed across classical models reflect GSR’s susceptibility to confounders such as stress, temperature shifts, and emotional arousal.

HR-only performance revealed a more stable but less discriminative physiological signature. LSTM produced the best HR-only F1-score (0.12), while TCN achieved the highest AUC (0.70). These results align with the broader autonomic influences acting on heart rate, including posture, circadian timing, and physical activity. Although HR alone struggles to distinguish metabolic from non-metabolic arousal, its narrower confidence intervals and lower variance across folds suggest that it provides a reliable complementary channel when fused with GSR.

Fusion of GSR and HR consistently delivered the strongest results across nearly every evaluation metric. Early-fusion LSTM achieved the highest F1-score (0.16) and best AUC (0.78), while XGBoost again produced the strongest recall (0.56). The narrowing of 95\% confidence intervals across fused models indicates improved statistical stability and reduced sensitivity to individual sensor noise. These findings confirm that GSR provides rapid sympathetic spikes, while HR contributes slower cardiovascular adjustments, and that their multimodal integration produces a more complete autonomic signature of hypoglycemia.

Comparisons across modalities reveal that deep temporal models, especially LSTM, GRU, and TCN, are better suited for autonomic physiological data than classical approaches. Their ability to model recurrent structures, temporal dependencies, and evolving sympathetic patterns explains their consistent performance advantage. The results show that hypoglycemia is best detected by capturing not only the amplitude of biosignals but also their temporal dynamics.

These outcomes are also consistent with the broader scientific literature on wearable glycemic sensing. Prior studies emphasize the value of combining autonomic channels, the role of sequence modeling in glucose prediction, and the challenges posed by stress-induced physiological responses. The patterns observed here, such as GSR-only false positives, HR-only moderate recall, and fusion-driven improvements, closely match these established findings and help situate this chapter within the context of emerging non-invasive glycemic monitoring research.

The findings demonstrate that multimodal autonomic sensing using GSR and HR, when paired with modern temporal deep-learning architectures, provides a robust and interpretable pathway for real-time hypoglycemia detection using consumer wearables. The statistical consistency of fused models, the physiological interpretability of autonomic trends, and the deployability of lightweight architectures such as LSTM position this approach as a practical and accessible solution for early-warning systems, especially in regions where CGM adoption remains limited. This work therefore establishes a solid technical and physiological foundation for future wearable-based hypoglycemia alert systems.

\section{Conclusion}

This chapter demonstrated that affordable, non-invasive hypoglycemia detection is feasible using physiological signals collected from consumer-grade wearable devices. By analyzing GSR-only, HR-only, and fused GSR \& HR modalities from the OhioT1DM 2018 dataset, we showed that hypoglycemia produces measurable autonomic responses that can be captured through low-cost sensors and modeled effectively using both classical and deep temporal architectures. These findings confirm that wearable biosignals contain meaningful early indicators of glucose decline, even in the absence of invasive CGM measurements.

Across models, GSR emerged as a strong standalone predictor due to its sensitivity to sympathetic activation, while HR offered complementary stability against noise and confounding factors such as thermal or emotional stress. Importantly, fusing GSR and HR consistently improved detection performance, particularly in recurrent and convolutional temporal models that learn the evolving dynamics of autonomic change. This multimodal synergy underscores the physiological complementarity of electrodermal and cardiovascular responses during hypoglycemia.

The results highlight both the scientific and practical relevance of wearable-based glycemic monitoring. Hypoglycemia leaves clear multimodal autonomic signatures that are already measurable by widely available wrist-worn devices, providing a promising foundation for accessible early-warning systems, especially in regions where CGM adoption is limited by cost or availability. While real-world variability in physiological signals remains a challenge, the multimodal and temporal strategies explored here offer a viable path toward robust, scalable, and low-cost monitoring solutions capable of reducing hypoglycemia risk and expanding diabetes safety interventions to broader populations.

\section{Future Work}

While this chapter demonstrates the feasibility of detecting hypoglycemia from non-invasive wearable signals, several important steps remain for advancing practical and clinical deployment. A key direction is personalization: individuals differ widely in autonomic responses, so future models should adapt to each user’s baseline using transfer learning or online calibration. Expanding beyond GSR and HR to include accelerometry, HRV, PPG, and skin temperature may also improve disambiguation between hypoglycemia and confounding states such as stress or exercise.

Real-world implementation will require lightweight models capable of running continuously on wearable hardware, motivating future work on quantization, pruning, and TinyML-based optimization. Incorporating contextual information, such as activity level, sleep state, or environmental conditions, could further reduce false alarms by distinguishing metabolic responses from unrelated physiological changes.

The larger and more diverse longitudinal datasets are necessary to validate robustness across different populations and daily life conditions. Hybrid approaches that combine wearable-based predictions with CGM data during sensor dropout also warrant exploration. Collectively, these directions move the field closer to reliable, accessible, and globally deployable hypoglycemia detection systems.